\documentclass[superscriptaddress,12pt]{revtex4-2} 
\usepackage[justification=justified,labelfont=bf,format=plain]{caption}
\DeclareCaptionFormat{myformat}{\fontsize{10}{10} \selectfont#1#2#3}
\usepackage{amsmath}  % needed for \tfrac, \bmatrix, etc.
\usepackage{amsfonts} % needed for bold Greek, Fraktur, and blackboard bold
\usepackage{graphicx} % needed for figures
\usepackage{esvect}
\usepackage{mathtools}
\usepackage{subcaption}
\usepackage{xcolor}
\usepackage{ulem}
\usepackage{appendix}
\usepackage{siunitx} 
\usepackage{placeins}

\usepackage{float} 

\usepackage{pbox}

\usepackage{mathrsfs}

\begin{document}

% Be sure to use the \title, \author, \affiliation, and \abstract macros
% to format your title page.  Don't use lower-level macros to  manually
% adjust the fonts and centering.

\title{Fractatomic Physics: An Invitation \\ with Atomic Stability and Rydberg States in Fractal Spaces}
% In a long title you can use \\ to force a line break at a certain location.

%When submitting the manuscript for review, do not include the author's name or institution

\author{Nhat A. Nghiem}
\affiliation{C. N. Yang Institute for Theoretical Physics, \\ State University of New York at Stony Brook, Stony Brook, NY 11794-3840, USA}
\affiliation{Department of Physics and Astronomy, \\ State University of New York at Stony Brook, Stony Brook, NY 11794-3800, USA.}

\author{Trung V. Phan}
\email{{tphan@natsci.claremont.edu}}
\affiliation{Department of Natural Sciences, Scripps and Pitzer Colleges, \\ Claremont Colleges Consortium, Claremont, CA 91711, USA}

\begin{abstract}

We explore the physical quantum properties of atoms in fractal spaces, both as a theoretical generalization of normal integer-dimensional Euclidean spaces and as an experimentally realizable setting. We identify the threshold of fractality at which Ehrenfest atomic instability emerges, where the Schr\"{o}dinger equation describing the wave-function of a single electron orbiting around an atom becomes scale-free, and discuss the potential of observing this phenomena in laboratory settings. We then study the Rydberg states of stable atoms using the Wentzel-Kramers-Brillouin approximation, along with a proposed extension for the Langer modification, in general fractal dimensionalities. We show that fractal space atoms near instability explode in size even at low-number excited state, making them highly suitable to induce strong entanglements and foster long-range many-body interactions. We argue that atomic physics in fractal spaces -- ``fractatomic physics'' -- is a rich research avenue deserving of further theoretical and experimental investigations. 
\end{abstract}

\date{\today}

\maketitle % title page is now complete

\section{Introduction}

\ \

Atomic physics lays the foundation for understanding the elementary constituents and basic interactions of matter \cite{kleppner1999short}. Perhaps one of the simplest yet most thought-provoking theoretical findings about atoms is that, in terms of classical orbital stability, they cannot remain stable in Euclidean spaces above four dimensions \cite{ehrenfest1917way,tangherlini1963schwarzschild,tegmark1997dimensionality}, contributing to many anthropic arguments for why our universe has only three \cite{pardo2018limits}. In laboratory settings, it is also possible to experimentally observe atomic instability around normal charges caused by massless (``relativistic'') quasi-electrons residing in two-dimensional lattice materials \cite{shytov2007atomic,casolo2009understanding,wang2013observing}. Exploring fundamentals in unconventional spaces can reveal new phenomena and offer unique insights into known physical principles \cite{parker1980one,pinto1993rydberg,nottale1996scale,maker1999quantum,kupriyanov2013hydrogen}, with vast potential for engineering applications and technologies \cite{allnatt2003atomic,datta2005quantum}. An interesting class of spaces to investigate is fractal, whose geometry exhibits self-similarity across scales \cite{mandelbrot1982fractal,barnsley2014fractals} and phenomena emerging within not only deviate significantly from their expected behaviors in conventional Euclidean spaces but also introduce new behaviors \cite{phan2020bacterial,martinez2022morphological,arellano2023habitat}. There have been many successes in generalizing and discovering novel physics by studying well-established models under fractality \cite{phan2024vanishing,tarasov2014flow,tarasov2014anisotropic,tarasov2015electromagnetic,tarasov2015elasticity,tarasov2015fractal,tarasov2016acoustic,naqvi2017electromagnetic}, using simple rules for vector calculus \cite{tarasov2011fractional,klafter2012fractional,tarasov2021general} founded on the assertion that mediums existing in fractal spaces can be effectively described using fractional continuous models \cite{tarasov2005continuous,tarasov2005possible}.

\ \ 

In this brief report, we investigate the quantum aspects of Hydrogen-like atoms in fractal spaces. Each atom is a bound-state of a stationary positively charged nucleus and a massive (``non-relativistic'') electron of negative charge. Here, we examine two distinct scenarios: one arises from mathematical curiosity, extending the concept of dimensionalities \cite{mattila1999geometry,tarasov2011fractional,fernandez2014fractal}, while the other considers experimental implementation within three-dimensional Euclidean space of our universe \cite{tarasov2013review}. To be more precise, in the former case (called {\it the full fractal space}), both the electrical field from the nucleus and the electron propagate within the same fractal space. In the later case (called {\it the embedded fractal space}), inspired by recent experiments for atomic collapse on graphene \cite{wang2013observing}, the electrical field spreads throughout all three dimensions, while the electron is a quasi-particle living on a fractal lattice embedded within those three dimensions. There have been studies on atomic physics in fractal spaces \cite{domany1983solutions,dong2008space,rodnianski2000fractal,chen2015spectral,liaqat2022novel,nottale1998scale} and intersecting spaces without defined dimensionalities \cite{he2023bound,ji1992quantum,sols1989theory}, but we believe we have offered a different perspective i.e. on how atomic structure can be manipulated with fractality.

\begin{figure*}[!htbp]
\includegraphics[width=0.8\textwidth]{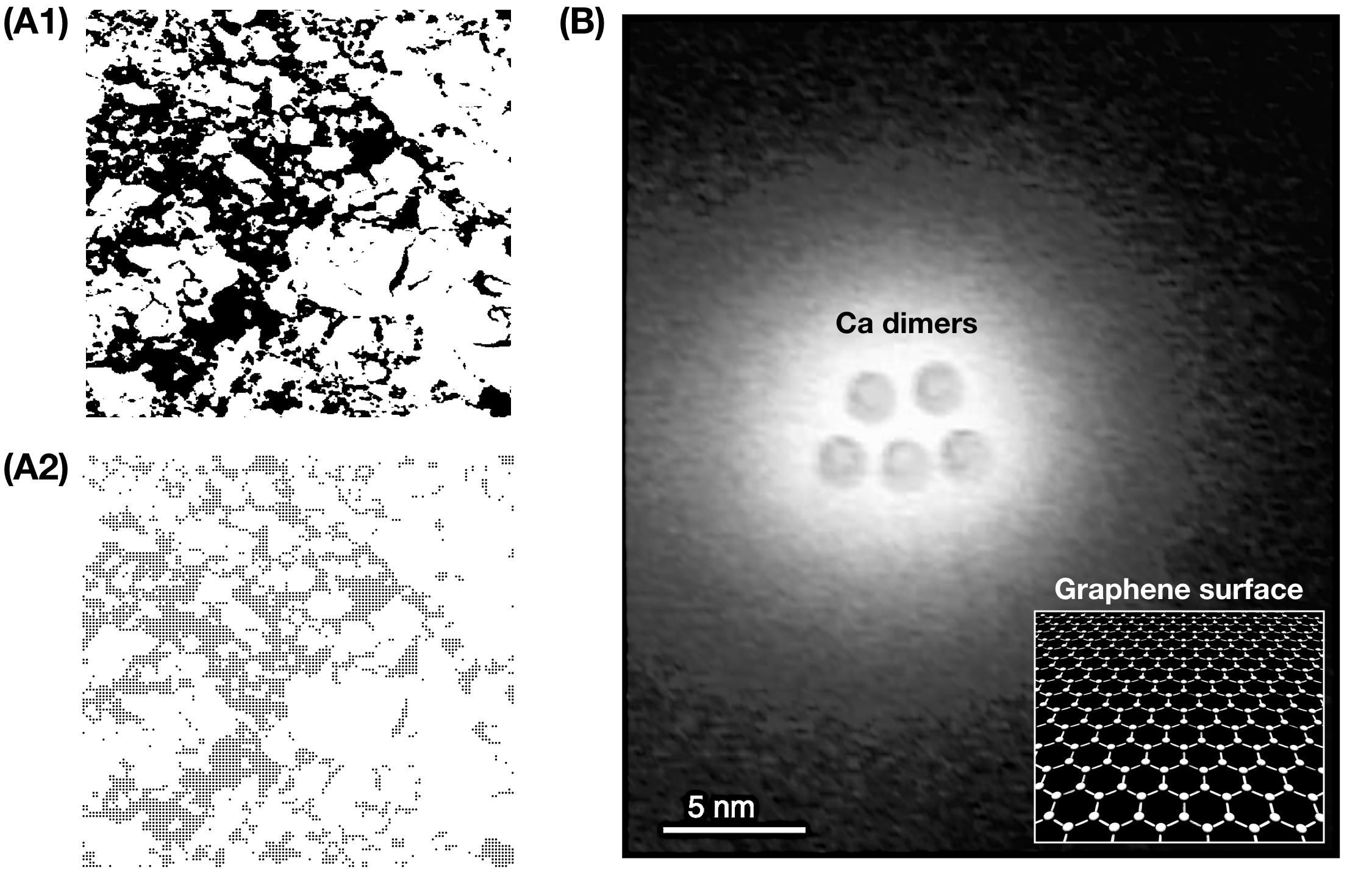}
\caption{\textbf{A fractal lattice and an artificial atom created in the lab.} {\bf (A1)} A two-dimensional slice of a fractal ``blueprint'' from normal three-dimensional natural soil, with fractality $\left( \mathscr{D}_{\text{v}},\mathscr{D}_{\text{v}}\right) = \left( 1.79,1.48 \right)$ \cite{phan2024vanishing,gimenez1997fractal}; the black area represents the open space. {\bf (A2)} The corresponding fractal lattice embedded in three-dimensional Euclidean space, in which dots are lattice sites and hopping between neighbor sites serves to establish an effective continuity within the material. {\bf (B)} An artificial nucleus -- made of five Calcium dimers placed on a two-dimensional graphene surface -- created in laboratory setting  \cite{wang2013observing}; the bright color representing the high $dI/d\phi$ signal (differential current with respect to electrical potential change) when probing the material.}
\label{fig01}
\end{figure*}

\ \ 

We begin by introducing a simple vector calculus formalism in fractal spaces. We briefly revisit the classical mechanics argument by Ehrenfest for why atoms should be unstable in higher-dimensional Euclidean spaces \cite{ehrenfest1917way}. Then, under a quantum mechanical perspective, we demonstrate that the instability condition is triggered by the emergence of a scale-free Schr\"{o}dinger equation describing the electron wavefunction around the nucleus \cite{shytov2007atomic}. This scale-free condition can be generalized for fractal spaces, allowing us to map out the boundary of fractality where the atomic transition from being stable to being unstable happens. We also find that there are macroscopic spatial structures in nature with fractional dimensions where atomic instability can occur, i.e. natural soil \cite{gimenez1997fractal,phan2024vanishing}. Thus, they might provide a ``blueprint'' (see Fig. \ref{fig01}A) for designing fractal lattices in the lab using nanoscale molecular engineering \cite{shang2015assembling,kempkes2019design,chen2015fractal}, through which we could search for the realization of Ehrenfest ``non-relativistic'' atomic instability phenomena \cite{ehrenfest1917way,tangherlini1963schwarzschild,tegmark1997dimensionality}. Note that we need the ground-state of these {\it embedded fractal space} atoms to be much larger compared to the lattice spacing, which can be controlled by the effective mass of the quasi-electron orbitting around the seeded nucleus \cite{wang2013observing} (see Fig. \ref{fig01}B). Finally, we investigate the physical properties of stable fractal space atoms by examining the behavior of their Rydberg (highly excited states \cite{gallagher1994rydberg,saffman2010quantum}) energy levels, using the Wentzel-Kramers-Brillouin (WKB) approximation \cite{wentzel1926verallgemeinerung,kramers1926wellenmechanik,brillouin1926mecanique,ao2023schrodinger} with our proposed extension for the Langer modification \cite{langer1937connection,gu2008improved}. We reveal that fractal space atoms exhibit a remarkable expansion in size as they approach the instability threshold, even at low-number excited states, as compared to that of their ground-state (which has been bounded by the lattice spacing for the fractality description to emerge). Consequently, these extremely gigantic atoms are effective at to triggering robust entanglements and facilitating long-range many-body interactions \cite{kleppner1981highly,gallagher1988rydberg}. These features might open up opportunities for advancing quantum technology \cite{adams2019rydberg} and proposing innovative applications, such as with quantum computing \cite{ryabtsev2005applicability}, quantum information \cite{saffman2010quantum}, quantum simulation \cite{weimer2010rydberg}, and even novel materials with exotic properties \cite{browaeys2020many}.

\ \ 

\section{Vector Calculus in Fractal Spaces}

\ \ 

Fractal spaces introduce unconventional physics that require a corresponding adjustment of mathematics. To be precise, the standard vector calculus applied for Euclidean spaces has to be generalized. Let us begin with some fractal geometric characteristics. Fractal spaces can be distinguished by their unique topological attributes \cite{weibel1991fractal,chen1998fractal}, in which the volume fractal dimension $\mathscr{D}_{\text{v}}$ and the surface fractal dimension $\mathscr{D}_{\text{s}}$ are the two most primary measures \cite{paumgartner1981resolution,friesen1987fractal,gimenez1997fractal}. Consider a spherical ball $\mathcal{S}$ of radius $r$ in this space, with surface $\partial \mathcal{S}$. The fractal dimensionalities $\text{dim}\left( \mathcal{S}\right) = \mathscr{D}_{\text{v}}$ and $\text{dim}\left( \partial \mathcal{S}\right) = \mathscr{D}_{\text{s}}$ can be positive non-integers. In general, the volume $\mathcal{V}(r)$ and the surface area of this ball $\mathcal{A}(r)$ follow power-law dependencies, i.e. $\mathcal{V}(r) \propto r^{\mathscr{D}_\text{v}}$ and $\mathcal{A}(r) \propto r^{\mathscr{D}_\text{s}}$, with the coefficients of proportionality depend on the particular spatial topology. For simplicity, we just assume these functions to be analytical continuations of those defined for hyperspheres \cite{das1990hyperspheres}:
\begin{equation}
\mathcal{V}(r) = \frac{\displaystyle \pi^{\mathscr{D}_{\text{v}}/2}}{\displaystyle \Gamma \left( \frac{\mathscr{D}_{\text{v}}+2}{2}\right)} r^{\mathscr{D}_{\text{v}}}  \ \ \text{and} \ \ \mathcal{A}(r) = \frac{\displaystyle 2\pi^{(\mathscr{D}_{\text{s}}+1)/2}}{\displaystyle \Gamma \left( \frac{\mathscr{D}_{\text{s}}+1}{2}\right)} r^{\mathscr{D}_{\text{s}}} \ \ ,
\label{fractal_volume_surface}
\end{equation}
in which $\Gamma(...)$ refers to the Gamma function, an extension of the factorial function to complex numbers \cite{edwards2001riemann}. Beside $\mathscr{D}_{\text{v}} > \mathscr{D}_{\text{s}}$, the constraint of embedding in the physical three-dimensional Euclidean space also bounds the fractal dimensions, i.e. $\mathscr{D}_{\text{v}}\leq 3$, $\mathscr{D}_{\text{s}} \leq 2$. 

\ \ 

We consider a Hydrogen-like atom, consisting of an atomic nucleus with a charge of $Z|q_{\text{e}}|$ ($Z \in \mathbb{N}$), which remains fixed in space, and an outer electron of charge $-|q_{\text{e}}|$ and mass $m_{\text{e}}$, located at a distance denoted as $r$ away from the nucleus. Here, we consider the electron of this Hydrogen-like atom to be in a s-orbital with a  spherically symmetric wave-functions, which should be a complex scalar function $\Psi(r)$. In this fractal space, The Laplacian operator $\Delta$ acting on any radial scalar function $S(r)$ gives \cite{phan2024vanishing}:
\begin{equation}
\Delta S(r) = F\left( \mathscr{D}_{\text{v}}, \mathscr{D}_{\text{s}} \right) \left\{ \left[ \frac1{r^{2\left( \mathscr{D}_{\text{v}} - \mathscr{D}_{\text{s}}\right) - 2}} \right] \partial_r^2 + \left[ \frac{-\mathscr{D}_{\text{v}} + 2 \mathscr{D}_{\text{s}} + 1}{r^{2\left( \mathscr{D}_{\text{v}} - \mathscr{D}_{\text{s}}\right) - 1}} \right] \partial_r \right\} S(r) \ \ ,
\label{fractal_laplacian}
\end{equation}
where the pre-factor is given by:
\begin{equation}
F\left( \mathscr{D}_{\text{v}}, \mathscr{D}_{\text{s}} \right) = \frac{\displaystyle \Gamma\left(\frac{\mathscr{D}_{\text{v}} - \mathscr{D}_{\text{s}}}{2}\right) \Gamma\left( \frac{\mathscr{D}_{\text{v}}}{2}\right)}{\displaystyle \pi^{\left( \mathscr{D}_{\text{v}} - \mathscr{D}_{\text{s}}\right)-1/2} \ \Gamma\left( \frac{\mathscr{D}_{\text{s}} +1}{2}\right)} \ \ .
\label{the_F}
\end{equation}
This result relies on an extension of the Green–Gauss theorem to fractal objects through fractional integrals in Euclidean spaces \cite{ostoja2009continuum,tarasov2014flow}. The Coulomb central interaction energy potential $U(r)$, felt by the electron due to the electrical field created by the nucleus, has the general form of a power-law:
\begin{equation}
    U(r) = - \text{sgn}(\kappa) \left| \mathbb{U} \right| r^{-\kappa} \ \ ,    \label{pot_gen}
\end{equation}
in which $\text{sgn}(\star)$ is the sign-function (equals to $+1$ when $\star >0$ and $-1$ when $\star <0$). We choose this expression for $U(r)$ so that it can be seen easily that this is an attractive potential. As mentioned in the introduction, we look into the following two scenarios:

\begin{itemize}
    \item {\it The full fractal space}, in which the electrical field and the electron propagate within the same fractal space. The interacting energy potential can be found to be:
    \begin{equation}
    \left| \mathbb{U}_{\text{ful}} \right| = \left| \frac{\displaystyle \Gamma \left( \frac{\mathscr{D}_{\text{s}}+1}{2}\right)}{\displaystyle 2 \kappa_{\text{ful}}  \pi^{(\kappa_{\text{ful}}+1)/2} 
 \ \Gamma \left( \frac{\mathscr{D}_{\text{v}}-\mathscr{D}_{\text{s}}}{2}\right)} \right| Z|q_{\text{e}}|^2 \ \ \text{and} \ \ \kappa_{\text{ful}} =2\mathscr{D}_{\text{s}} - \mathscr{D}_{\text{v}}  \ \ .
    \label{pot_frac}
    \end{equation}
    We show the derivation of this using fractal vector calculus in Appendix \ref{Cou_frac}. 
    %The exponent $\kappa_{\text{ful}}$ changes sign at $\mathscr{D}_{\text{v}}/\mathscr{D}_{\text{s}}=1/2$, causes $U(r)$ to no longer obey a power-law ($\kappa_{\text{ful}}=0$) but a logarithmic function.
    \item {\it The embedded fractal space}, where the electrical field spreads throughout the three dimensional Euclidean space of our universe while the electron is a quasi-particle residing on a fractal lattice embedded within those three dimensions. The interacting energy potential has the familiar form:
    \begin{equation}
        \left| \mathbb{U}_{\text{emb}} \right| = \left( \frac{1}{4\pi} \right) Z|q_{\text{e}}|^2 \ \ \text{and} \ \ \kappa_{\text{emb}}=1  \ \ .
    \label{pot_norm}
    \end{equation}
    This can also be found by setting $\left(\mathscr{D}_{\text{v}},\mathscr{D}_{\text{s}}\right)=\left(3,2\right)$ in Eq. \eqref{pot_frac}.
\end{itemize}
The former embarks on a mathematical adventure, while the latter is motivated by potential realizations within laboratory settings.

\ \ 

We emphasize that the assumption underlying all these mathematics -- fractional continuous models for mediums on fractal spaces \cite{tarasov2005continuous,tarasov2005possible} -- should not be viewed as definitive but as a heuristic tool offering glimpses into the physics in fractal spaces \cite{phan2024vanishing}.

\ \ 

\section{Atomic Instability}

\ \ 

We investigate the dimensionalities of space at which atoms become unstable, beginning with revisiting classical mechanics in Euclidean spaces, followed by expanding this argument into a quantum mechanical perspective in Euclidean spaces, and concluding with a complete and generalized model to describe atomic behaviors in fractal spaces. Note that, $\mathscr{D}$-dimensional Euclidean spaces are special fractal spaces with volume and surface dimensionalities are given by $\left(\mathscr{D}_{\text{v}},\mathscr{D}_{\text{s}} \right) = \left(\mathscr{D},\mathscr{D}-1\right)$, in which $\mathscr{D}$ is a natural number i.e. $\mathscr{D} \in \mathbb{N}$. 

\ \ 

\subsection{Ehrenfest Finding in Euclidean Spaces \label{E_arg}}

\ \

In classical mechanics, the planetary model depicts Hydrogen-like atoms with an electron orbiting around the nucleus in planar \cite{rutherford2012scattering,bohr1921atomic,lakhtakia1996models}. The electrical potential $U(r)$ created around the nucleus is spherical symmetric, therefore angular momentum $L$ is conserved. Consider the radial motion $r(t)$ of the electron ($t$ represents time), we can define an effective potential $U_{\text{eff}}(r)$ in which the radial acceleration $\ddot{r} = m_{\text{e}}^{-1} \partial_r U_{\text{eff}}(r)$:
\begin{equation}
U_{\text{eff}}(r) = U(r) + \frac{L^2}{2m_{\text{e}}r^2}
\label{pot_eff}
\end{equation}
The extra energy-term $L^2/2m_{\text{e}}r^2$ is the kinetic energy stored inside the compact angular dimension \cite{phan2021curious}. Following from Eq. \eqref{pot_frac}, in $\mathscr{D}$-dimensional Euclidean spaces:
\begin{equation}
\kappa_{\text{ful}} = \mathscr{D}-2 \ \ \text{and therefore} \ \ U(r) \propto -r^{-\left(\mathscr{D}-2 \right)} \ \ ,
\label{pot_Euc}
\end{equation}
which means $U_{\text{eff}}(r)$ in Eq. \eqref{pot_eff} is monotonic for $\mathscr{D} \geq 4$ and there is no minimum except at $r=0$ and $r=+\infty$ (corresponding to atomic collapse and atomic deconfinement). No stable orbit for an electron around the nucleus, hence classically atom cannot be stable in Euclidean spaces with dimension $\mathscr{D}=4$ and above.

\ \ 

\subsection{Scale-Free Schr\"{o}dinger Equation in Euclidean Spaces \label{sec:Schr_scalefree}}

\ \ 

Going from classical to quantum mechanics \cite{griffiths1984consistent,omnes1992consistent,zurek2002decoherence,joos2013decoherence}, the stationary state of atom can be described by the time-independent Sch\"{o}dinger equation:
\begin{equation}
\hat{H} \Psi(r) = E \Psi(r) \ \ , \ \ \hat{H} = -\frac{\hbar^2}{2m_{\text{e}}} \Delta + U(r) \ \ ,
\label{Schr_eq}
\end{equation}
in which $\Psi(r)$, $m_e$, and $E$ are the wavefunction, the mass, and the corresponding energy of the outer electron around the nucleus. $\hat{H}$ is the Hamiltonian operator, $\hbar$ is the reduced Planck constant, and $\Delta$ is the Laplacian operator. The Sch\"{o}dinger equation becomes scale-free when an homogeneous rescaling of $r \rightarrow \gamma r$ leads to a rescaling of $\hat{H} \rightarrow \gamma' \hat{H}$. When this occurs, the atomic spectrum becomes continuous. This is because if the wavefunction $\Psi(r)$ is a solution of Eq. \eqref{Schr_eq} with energy $E$, then the rescaled wavefunction $\Psi(\gamma r)$ with rescaled energy $\gamma' E$ is also a solution. If the bound state requires $E < 0$, then the possible energy for bound states is unbounded from below (the energy can be arbitrarily negatively large, indicating no well-defined ground-state and thus atomic instability).

\ \ 

For the scale-free condition to be satisfied, the Laplacian $\Delta$ and the power-law energy potential $U(r)$ must have the same spatial-scaling. In Euclidean spaces, $\Delta \propto r^{-2}$ always (as a second-derivative in space), hence we get independent of scale when $U(r)\propto r^{-2}$. From Eq. \eqref{pot_Euc}, we know that $U(r)\propto r^{-(\mathscr{D}-2)}$, hence the scale-free critical dimension should be at $\mathscr{D}-2=2$ i.e. $\mathscr{D}=4$. This is also the smallest dimensionality of Euclidean spaces where Ehrenfest atomic instability is expected to emerge in classical mechanics, as explained in Section \ref{E_arg}.

\ \ 

\subsection{Scale-Free Schr\"{o}dinger Equation in Fractal Spaces \label{sec:Schr_scalefree_frac}}

\ \

In a general fractal spaces, the Laplacian operator scales differently. From Eq. \eqref{fractal_laplacian}, we have the scaling rule $\Delta \propto r^{-2\left( \mathscr{D}_{\text{v}} - \mathscr{D}_{\text{s}} \right)}$. For a power-law energy potential $U(r)$ as in Eq. \eqref{pot_gen}, the scale-free condition emerges when the range of interaction $\kappa$ satisfies:
\begin{equation}
\kappa = 2\left( \mathscr{D}_{\text{v}} - \mathscr{D}_{\text{s}} \right) \ \ .
\label{stab_cond}
\end{equation}
Let us look at this closer in two scenarios of interests:
\begin{itemize}
    \item {\it The full fractal space} has $\kappa_{\text{ful}}=2\mathscr{D}_{\text{s}}- \mathscr{D}_{\text{v}}$, as found in Eq. \eqref{pot_frac}. Therefore, the fractality for scale-free atom is:
    \begin{equation}
    \frac{\mathscr{D}_{\text{v}}}{\mathscr{D}_{\text{s}}} = \frac{4}{3} \ \ .
    \label{scalefree_frac}
    \end{equation}
    This agrees with what we have found in Section \ref{sec:Schr_scalefree} for Euclidean spaces. 
    \item {\it The embedded fractal space} has $\kappa_{\text{emb}}=1$, see Eq. \eqref{pot_norm}. Thus, scale-free atom can be found at fractality satisfies:
    \begin{equation}
     \mathscr{D}_{\text{v}} - \mathscr{D}_{\text{s}}  = \frac12 \ \ .
    \label{scalefree_norm}
    \end{equation}
    This indicates that the radial dimension should exhibits a strong fractal characteristics.
\end{itemize}
With Eq. \eqref{scalefree_frac} or Eq. \eqref{scalefree_norm} as the transition threshold in dimensionality-space $\left( \mathscr{D}_{\text{v}}, \mathscr{D}_{\text{s}} \right)$, we realize that the fractality found in natural soil (see Fig. \ref{fig01}A) can give us unstable atoms in both cases, e.g. $\left( \mathscr{D}_{\text{v}}, \mathscr{D}_{\text{s}} \right) = \left( 1.79,1.48\right)$ \cite{gimenez1997fractal,phan2024vanishing}. 

\ \ 

\section{Rydberg States}

\ \ 

In Section \ref{sec:Schr_scalefree_frac}, we have found that atoms are stable in $\mathscr{D}_{\text{v}}/ \mathscr{D}_{\text{s}} > 4/3$ for {\it the full fractal space} and $\mathscr{D}_{\text{v}}- \mathscr{D}_{\text{s}} > 1/2$ for {\it the embedded fractal space}. We can further explore some quantum mechanical properties of these stable fractal space atoms. Let us demonstrate that by investigating the Rydberg states of Hydrogen-like atoms. To be precise, we study how their excited state energy-level $E(n)$ depends on  their large quantum number $n \gg 1$, $n \in \mathbb{N}$ \cite{gallagher1994rydberg,saffman2010quantum}. Note that this model aims to understand the overall scaling of atomic energy without considering special contributions e.g. from quantum defects \cite{Lorenzen1983}.

\ \ 

We define the rescaled radius and the rescaled energy as:
\begin{equation}
\tilde{r} = r \left[ \frac{\hbar^2 F\left( \mathcal{D}_{\text{v}},\mathcal{D}_{\text{s}} \right)}{m_{\text{e}} \left| \mathbb{U}\right|} \right]^{\frac{1}{\kappa-2 \left( \mathcal{D}_{\text{v}}-\mathcal{D}_{\text{s}} \right)}} \ \ , \ \ \tilde{E} =  E \left| \mathbb{U} \right|^{-1} \left[ \frac{m_{\text{e}} \left| \mathbb{U} \right|}{\hbar^2 F\left( \mathcal{D}_{\text{v}},\mathcal{D}_{\text{s}} \right)} \right]^{\frac{ \kappa  }{\kappa-2 \left( \mathcal{D}_{\text{v}}-\mathcal{D}_{\text{s}} \right) }} \ \ ,
\end{equation}
in which $F\left( \mathscr{D}_{\text{v}},\mathscr{D}_{\text{s}} \right)$ is given by Eq. \eqref{the_F}. We rewrite the Schr\"{o}dinger equation from Eq. \eqref{Schr_eq}, using the fractal Laplacian operator in Eq. \eqref{fractal_laplacian} and the power-law potential $U(r)$ in Eq. \eqref{pot_gen}, as follows:
\begin{equation}
\left\{ \partial_{\tilde{r}}^2 + \left[ \frac{-\mathscr{D}_{\text{v}} + 2\mathscr{D}_{\text{s}}+1}{\tilde{r}} \right] \partial_{\tilde{r}} + 2\tilde{r}^{2\left( \mathcal{D}_{\text{v}}-\mathcal{D}_{\text{s}}\right)-2} \left[ \tilde{E} + \text{sgn}(\kappa) \tilde{r}^{-\kappa} \right] \right\} \Psi(\tilde{r}) = 0 \ \ .
\end{equation}
We then define the rescaled wavefunction
\begin{equation}
\tilde{\Psi}(\tilde{r}) = \tilde{r}^{\frac{-\mathscr{D}_{\text{v}} + 2\mathscr{D}_{\text{s}}+1}2} \Psi(\tilde{r}) 
\end{equation}
to even further simplify the Schr\"{o}dinger equation:
\begin{equation}
\left( \partial_{\tilde{r}}^2 + \left\{ 2\tilde{r}^{2\left( \mathscr{D}_{\text{v}}-\mathscr{D}_{\text{s}}\right)-2} \left[ \tilde{E} + \text{sgn}(\kappa) \tilde{r}^{-\kappa} \right] - \left[ \frac{\left( -\mathscr{D}_{\text{v}} + 2 \mathscr{D}_{\text{s}}\right)^2-1}4\right] \tilde{r}^{-2} \right\} \right) \tilde{\Psi}(\tilde{r}) = 0 \ \ . 
\label{radial_Schr_resc}
\end{equation}

\ \

For analytical tractability, instead of solving Eq. \eqref{radial_Schr_resc} directly, we use WKB approximation \cite{wentzel1926verallgemeinerung,kramers1926wellenmechanik,brillouin1926mecanique,ao2023schrodinger}. The estimated radial momentum of the electron can be found by replacing $\partial_{\tilde{r}}$ with $i\tilde{p}_r(\tilde{r})$ in Eq. \eqref{radial_Schr_resc}:
\begin{equation}
\tilde{p}_r(\tilde{r}) = \left\{ 2\tilde{r}^{2\left( \mathscr{D}_{\text{v}}-\mathscr{D}_{\text{s}}\right)-2} \left[ \tilde{E} + \text{sgn}(\kappa) \tilde{r}^{-\kappa} \right] - \left[ \frac{\left( -\mathscr{D}_{\text{v}} + 2 \mathscr{D}_{\text{s}}\right)^2-1}4\right] \tilde{r}^{-2} \right\}^{1/2} \ \ .
\label{rad_mom}
\end{equation}
It is important to mention that the WKB approximation  has long been known to be problematic in general dimensions. To address this, a trick like the generalized Langer modification is often used for more accurate estimations of eigenenergies \cite{langer1937connection,gu2008improved}. Here, we follow the simple proposal made in \cite{gu2008improved}, applying it for spherical-mode of the wavefunction and extending its applicability to fractal spaces. We modify the radial momentum in Eq. \eqref{rad_mom} to:
\begin{equation}
\tilde{p}_r(\tilde{r}) \ \xrightarrow{\text{modify}} \  \tilde{p}'_r(\tilde{r}) = \left\{ 2\tilde{r}^{2\left( \mathscr{D}_{\text{v}}-\mathscr{D}_{\text{s}}\right)-2} \left[ \tilde{E} + \text{sgn}(\kappa) \tilde{r}^{-\kappa} \right] - \left[ \frac{\left( -\mathscr{D}_{\text{v}} + 2 \mathscr{D}_{\text{s}}\right)^2}4\right] \tilde{r}^{-2} \right\}^{1/2} \ \ ,
\label{rad_mom_mod}
\end{equation}
which can ensure that the estimations for the energy spectrums $\tilde{E}(n)$ in Euclidean spaces of $\mathscr{D} \in \mathbb{N}$ dimensions are reasonably good. This change has the benefit of being simple, yet rather far from being trivial, so we provide a detailed explanation in Appendix \ref{app:Langer_mod}. To get exactly correct eigenstate energies, we need to also change the Maslov index $\mu=2 \ \xrightarrow{\text{modify}} \ \mu'$ in a very complicated manner \cite{gu2008improved,watson1989semiclassical}.

\ \ 

\begin{figure*}[!htbp]
\includegraphics[width=\textwidth]{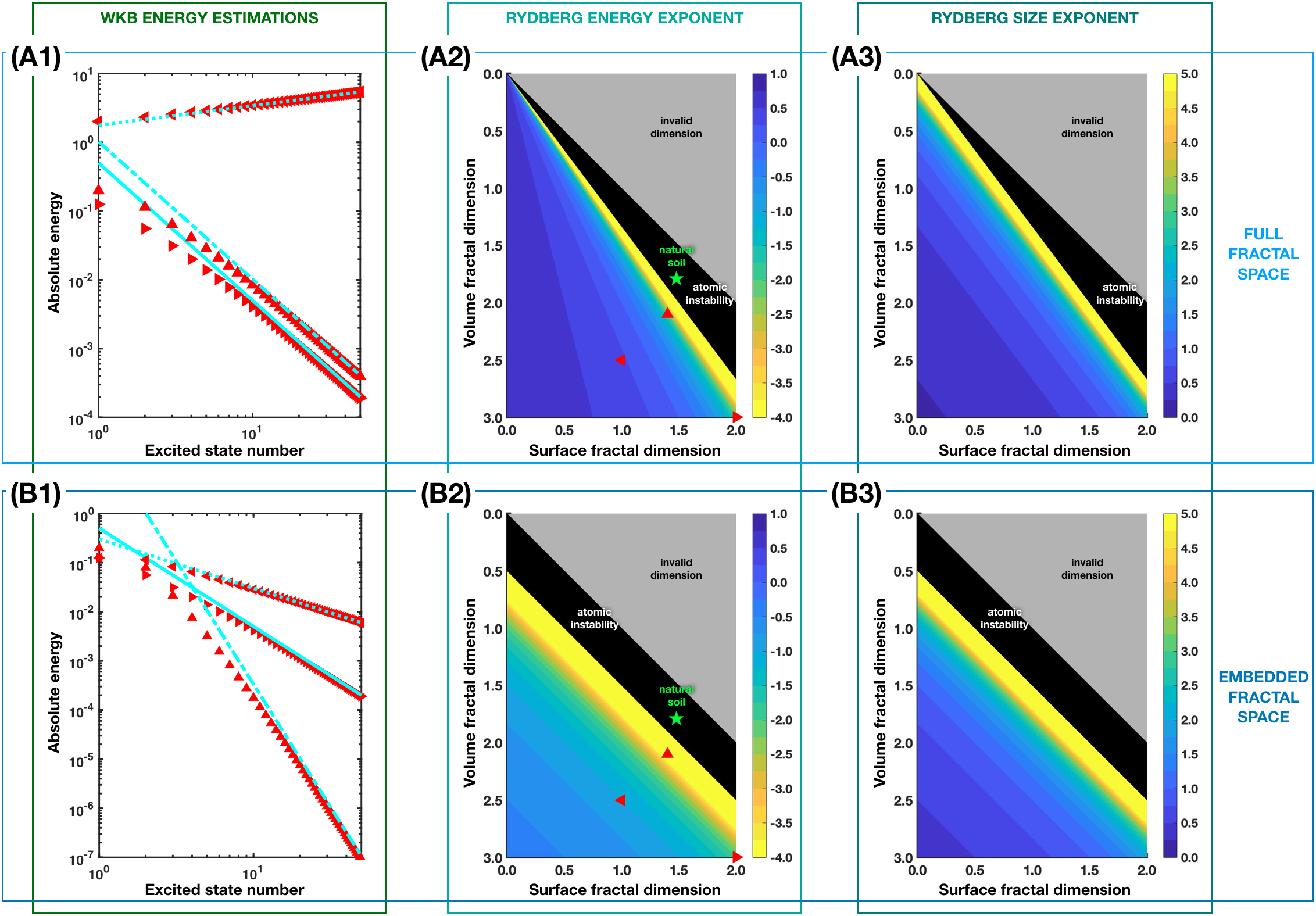}
\caption{\textbf{Numerical investigations of atomic properties at different fractalities.} The row {\bf (A)} are figures from {\it the full fractal space} scenario. {\bf (A1)} The estimated energy $| \tilde{E}|$ as a function of excited state number $n$ (from ground state $n=1$ to $n=50$) in log-log scale, following from solving Eq. \eqref{WS_quant} numerically. We then compare them with their expected asymptotic behavior, given by Eq. \eqref{R_E_asymp}. The fractal dimensionalities $\left( \mathscr{D}_{\text{v}},\mathscr{D}_{\text{s}} \right)$ we choose to investigate are $\left(3,2 \right)$, $\left(2.5,1 \right)$, and $\left(2.1,1.4 \right)$; their numerical values are given by triangle markers pointing right, pointing left, and pointing up, while their theoretical asymptotic behaviors are shown with continuous, dot-dot, and dot-dash lines. {\bf (A2)} Rydberg energy exponent and {\bf (A3)} Rydberg size exponent as a function of fractalities, from Eq. \eqref{Ryd_gen}. The row {\bf (B)} are figures from {\it the embedded fractal space} scenario, in which {\bf (B1-3)} are of similar descriptions with {\bf (A1-3)}.}
\label{fig02}
\end{figure*}

We have mentioned how the WKB approximation can be used to estimate the eigenstate energies, but we have not yet explained the method. Let us do that now. To find $\tilde{E}(n)$, we choose the ansatz
\begin{equation}
\tilde{E} = -\text{sgn}(\kappa) \left| \tilde{E} \right| 
\label{ansatz_E}
\end{equation}
and impose the Wilson-Sommerfeld quantization condition \cite{brack2018semiclassical}:
\begin{equation}
\frac12 \oint \tilde{p}'_r(\tilde{r}) d\tilde{r} = \int^{\tilde{r}_{\max}}_{\tilde{r}_{\min}} \tilde{p}'_r(\tilde{r}) d\tilde{r} = \pi \left[ (n-1) + \frac{\mu}4 \right] \ \ \text{in which} \ \ \tilde{p}'_r(\tilde{r}_{\min})=\tilde{p}'_r(\tilde{r}_{\max})=0 \ \ . 
\label{WS_quant}
\end{equation}
Note that $n=1$ corresponds to the ground-state in this counting convention. For simplicity, we keep using the Maslov index $\mu=2$. We numerically investigate the solution $\left|\tilde{E}\right|(n)$ and the corresponding $\tilde{r}_{\max}$ of Eq. \eqref{WS_quant} in Fig. \ref{fig02}A1 and Fig. \ref{fig02}B1 for some representative fractalities.

\ \  

Inside the square-root of the modified radial momentum, which is given by Eq. \eqref{rad_mom}, there are three radial-dependent terms with exponents $2\left( \mathscr{D}_{\text{v}}-\mathscr{D}_{\text{s}} \right)-2$, $2\left( \mathscr{D}_{\text{v}}-\mathscr{D}_{\text{s}} \right)-2-\kappa$, and $-2$. Since $\mathscr{D}_{v} > \mathscr{D}_{s}$ and the atomic stability condition in Eq. \eqref{stab_cond} requires $\kappa < 2\left( \mathscr{D}_{\text{v}}-\mathscr{D}_{\text{s}} \right)$, the third exponent should always be the smallest. For $n \gg 1$, the atomic size becomes large, then following from Eq. \eqref{rad_mom_mod}, Eq. \eqref{ansatz_E}, and Eq. \eqref{WS_quant} we can use the approximation:
\begin{equation}
\int^{\tilde{r}_{\max}}_{0} \left[ 2\tilde{r}^{2\left( \mathscr{D}_{\text{v}}-\mathscr{D}_{\text{s}}\right)-2} \ \text{sgn}(\kappa) \left( -\left|\tilde{E}\right| + \tilde{r}^{-\kappa} \right)  \right]^{1/2} d\tilde{r} \approx \pi n  \ \ \text{where} \ \ \tilde{r}_{\max} \approx \left|\tilde{E}\right|^{-1/\kappa} \ \ .
\label{wkb_int}
\end{equation}
This approximation is unaffected by how the Maslov index $\mu$ is chosen. We estimate the scaling of energy levels and sizes for fractal space atom Rydberg states to be:
\begin{equation}
\left| \tilde{E} \right| \propto n^{-\frac{2 \kappa}{2\left(\mathscr{D}_{\text{v}}-\mathscr{D}_{\text{s}} \right)-\kappa}} \ \ , \ \ \tilde{r}_{\max} \propto n^{\frac{2}{2\left(\mathscr{D}_{\text{v}}-\mathscr{D}_{\text{s}} \right)-\kappa}} \ \ .
\label{Ryd_gen}
\end{equation}
We show their derivation in Appendix \ref{app:WKB}. Let us look at them closer, in each of the two scenarios of interests:
\begin{itemize}
    \item {\it The full fractal space} has $\kappa_{\text{ful}}=2\mathscr{D}_{\text{s}}- \mathscr{D}_{\text{v}}$, as previously found in Eq. \eqref{pot_frac}. Thus:
    \begin{equation}
\left| \tilde{E} \right| \propto n^{-\frac{ 2 \left(2 \mathscr{D}_{\text{s}}-\mathscr{D}_{\text{v}} \right) }{3\mathscr{D}_{\text{v}}- 4 \mathscr{D}_{\text{s}}}} \ \ , \ \ \tilde{r}_{\max} \propto n^{\frac{2}{3\mathscr{D}_{\text{v}}- 4 \mathscr{D}_{\text{s}}}} \ \ .
\label{Ryd_frac}
\end{equation} 
    \item {\it The embedded fractal space} has $\kappa_{\text{emb}}=1$, from Eq. \eqref{pot_norm}. Therefore:
    \begin{equation}
     \left| \tilde{E} \right| \propto n^{-\frac{2}{2\left(\mathscr{D}_{\text{v}}-\mathscr{D}_{\text{s}} \right)-1}} \ \ , \ \ \tilde{r}_{\max} \propto n^{\frac{2}{2\left(\mathscr{D}_{\text{v}}-\mathscr{D}_{\text{s}} \right)-1}} \ \ .
     \label{Ryd_emb}
    \end{equation}
\end{itemize}
 These results are also in agreement with our representative numerical findings shown in Fig. \ref{fig02}A1 and Fig. \ref{fig02}B1. We plot them as surface functions of fractality $\left(\mathscr{D}_{\text{v}} , \mathscr{D}_{\text{s}}\right)$, for {\it the full fractal space} in Fig. \ref{fig02}A2 and Fig. \ref{fig02}A3, also for {\it the embedded fractal space} in Fig. \ref{fig02}B2 and Fig. \ref{fig02}B3. 

\ \

In the vicinity of the scale-free threshold, i.e. $\kappa \rightarrow 2\left(\mathscr{D}_{\text{v}}-\mathscr{D}_{\text{s}} \right)$ from below as found in Eq. \eqref{stab_cond}, the atomic size explodes after the stable fractal space atoms are excited from the ground state size, which follows directly from Eq. \eqref{Ryd_gen} as $\tilde{r}_{\max} \propto n^{1/0^+}$. For comparison, the usual case of a Rydberg atom in $\mathscr{D}=3$-dimensional Euclidean space has the atomic size scales as $\propto n^{2}$ \cite{gallagher1994rydberg}. This ability to drastically bloat up even at low excited states can help generating long-distance interactions in cold neutral atom systems. This allows atoms to interact on a macroscopic scale, facilitating long-distance quantum communication \cite{kimble2008quantum}.  Additionally, the strength of the Rydberg-Rydberg interactions will increase greatly, which is crucial for technologies such as quantum computing \cite{SaffmanBlockadeGate, DeutschDressedQC,CiracZollerLukin} and quantum simulation \cite{TurnerRydbergScar, DeutschMicroMacro,LewisSwanBoseHubbard}. With stronger interactions, entanglement between atoms occurs rapidly, enabling a significantly higher number of operations within a time that is negligible to the atoms decoherence time. In quantum simulation, the fast growth of entanglement enables us to study system dynamics on a longer timescale. This capability holds promise in gaining insight into systems exhibiting glass-like dynamics and slow thermalization (many-body localization regime \cite{abanin2019colloquium}), which are typically challenging to observe in traditional laboratory settings due to thermalization occurring at an inaccessible time.

\ \ 

\section{Discussion}

\ \ 

In this brief report, we give some remarks about atomic physics in fractal spaces. Beyond being merely a mathematical curiosity, we have demonstrated the potential in laboratory settings to observe Ehrenfest atomic instability and approach the scale-free boundary of fractality to create atoms that explode after becoming excited. These exotic properties of fractal space atom can be of relevant to the development of quantum engineering \cite{zagoskin2011quantum}. Here we have only scratched the surface of a deep topic that requires further research attempts. On a single atom level, we have yet to fully comprehend the intricacies of angular modes beyond spherical configurations \cite{varshalovich1988quantum,edmonds1996angular,fickler2012quantum}, scattering \cite{wu2014quantum,chadan2012inverse} and resonances \cite{moiseyev1998quantum}, and we also do not consider atoms possessing many electrons \cite{sucher1980foundations}. On a many-atom level, it remains largely untouched, leaving a vast uncharted territory awaiting exploration \cite{cirac1994preparation,heine1991many,kohn1995density}. We hope our work can ignite some interests in fractal space atoms, not just as a theoretical adventure, but also as an experimental quest that one can realisitcally pursue.

\ \ 

\section{Acknowledgement}

\ \

Nhat A. Nghiem acknowledges the support by a Seed Grant from
Stony Brook University’s Office of the Vice President for Research and by the Center for Distributed Quantum Processing. We would like to thank Van H. Do for many useful discussions.

\ \ 

\textbf{Declaration Statement:} We confirm that the manuscript has been read and approved by all named authors and that there are no other persons who satisfied the criteria for authorship but are not listed. We further confirm that the order of authors listed in the manuscript has been approved by all of us. There are no known conflicts of interest associated with this publication and there has been no significant financial support for this work that could have influenced its outcome.    

\ \

\appendix 

\ \ 

\section{The Coulomb Potential in Fractal Spaces \label{Cou_frac}}

\ \ 

For {\it the full fractal space}, from Gauss flux theorem \cite{singh2006student}, we have the radial electrical field $\vec{\mathcal{E}}(r) \propto \hat{r}$ surrounding the nucleus to be:
\begin{equation}
\oint_{\partial \mathcal{S}} d\vec{\sigma} \vec{\mathcal{E}}(r) = Z|q_{\text{e}}|  \ \ \Longrightarrow \ \ \vec{\mathcal{E}}(r) = \frac{Z|q_{\text{e}}|}{\mathcal{A}(r)} \hat{r} \ \ ,
\label{elec_field}
\end{equation}
where $\mathcal{A}(r)$ is given in Eq. \eqref{fractal_volume_surface}, $d\vec{\sigma} = d\sigma \hat{r} $ is the differential area normal-vector on the spherical surface $\partial \mathcal{S}$, and $\hat{r}$ is the radial unit vector. The electrical field $\vec{\mathcal{E}}$ is the gradient of the electrical potential $\phi(r)$, i.e. $\vec{\mathcal{E}} = \vec{\nabla} \phi(r)$; the $\vec{\nabla}$ operator in this fractal space acting on any radial scalar function $S(r)$ gives \cite{phan2024vanishing}:
\begin{equation}
\vec{\nabla} S(r) = \left[ \frac{\displaystyle \Gamma\left(\frac{ \mathscr{D}_{\text{v}} - \mathscr{D}_{\text{s}}}{2}\right)}{\displaystyle \pi^{\left( \mathscr{D}_{\text{v}} - \mathscr{D}_{\text{s}}\right)/2}} \frac1{\displaystyle r^{\left( \mathscr{D}_{\text{v}} - \mathscr{D}_{\text{s}}\right)-1}} \right] \partial_r S(r) \hat{r} \ \ .
\label{fractal_grad}
\end{equation}
In order to solve for $\phi(r)$, we consider the ansatz $\phi(r) = \Phi r^{-\kappa}$ where $\Phi$ and $\kappa$ are some constants to be determined. Using Eq. \eqref{fractal_volume_surface}, Eq. \eqref{elec_field}, and Eq. \eqref{fractal_grad}, with this ansatz we arrive at:
\begin{equation}
\kappa = 2\mathscr{D}_{\text{s}} - \mathscr{D}_{\text{v}} \ \ \text{and} \ \ \Phi = \left[ \frac{\displaystyle \Gamma \left( \frac{\mathscr{D}_{\text{s}}+1}{2}\right)}{\displaystyle 2 \kappa \pi^{(\kappa+1)/2} \ \Gamma \left( \frac{\mathscr{D}_{\text{v}}-\mathscr{D}_{\text{s}}}{2}\right)} \right]  Z|q_{\text{e}}| \ \ .
\end{equation}
Note that $\Phi$ and $\kappa$ always have the same sign, thus $\Phi = \text{sgn}(\kappa) |\Phi|$. The Coulomb central interacting energy potential is given by $U(r) = -|q_{\text{e}}|\phi(r)$.

\ \ 

\section{Generalized Langer Modification in Fractal Spaces \label{app:Langer_mod}}

\ \ 

The generalized Langer modification in Euclidean space of dimension $\mathscr{D}$ for wavefunction of angular mode $l$ is given by Eq. (24) in \cite{gu2008improved}, in which the pre-factor of the $r^{-2}$-terms inside the $\{...\}^{1/2}$ of Eq. \eqref{rad_mom}:
\begin{equation}
\frac{\left(\mathscr{D} + 2l - 2 \right)^2 - 1}4 \  \xrightarrow{\text{modify}} \ \frac{\left(\mathscr{D} + 2l - 2 \right)^2}4 \ \ ,
\label{langer_mod_gu}
\end{equation}
which after imposing the Wilson-Sommerfeld quantization condition \cite{brack2018semiclassical} and change the Maslov index $\mu=2 \ \xrightarrow{\text{modify}} \ \mu'$ makes the eigenstate energies correct \cite{gu2008improved,watson1989semiclassical}. For the spherical mode $l=0$ and at fractality $\left( \mathscr{D}_{\text{v}},\mathscr{D}_{\text{s}} \right)$ of space to be Euclidean $\left( \mathscr{D},\mathscr{D}-1 \right)$, this change in Eq. \eqref{langer_mod_gu} is the same as:
\begin{equation}
\frac{\left(-\mathscr{D}_{\text{v}} + 2\mathscr{D}_{\text{s}} \right)^2 - 1}4 \  \xrightarrow{\text{modify}} \ \frac{\left(-\mathscr{D}_{\text{v}} + 2\mathscr{D}_{\text{s}} \right)^2}4 \ \ ,
\end{equation} 
and this is indeed what we have done in Eq. \eqref{rad_mom_mod}.

\ \ 

\section{WKB Estimations for Rydberg Energies and Radius of Fractal Space Atoms \label{app:WKB}}

\ \ 

Define $z=\tilde{r}/\tilde{r}_{\max}$, we rewrite Eq. \eqref{wkb_int} as:
\begin{equation}
\tilde{r}_{\max}^{\left( \mathscr{D}_{\text{v}}-\mathscr{D}_{\text{s}}\right) -\kappa/2 } \int^{1}_{0} \left[ 2z^{2\left( \mathscr{D}_{\text{v}}-\mathscr{D}_{\text{s}}\right)-2} \ \text{sgn}(\kappa) \left( -1 + z^{-\kappa} \right)  \right]^{1/2} dz \approx \pi n  \ \ \text{and} \ \ \left|\tilde{E}\right| \approx  \tilde{r}_{\max}^{-\kappa} \ \ .
\label{wkb_int_dimless}
\end{equation}
The integral can be evaluated to be:
\begin{equation}
\begin{split}
\Theta = \int^{1}_{0} \left[ 2z^{2\left( \mathscr{D}_{\text{v}}-\mathscr{D}_{\text{s}}\right)-2} \ \text{sgn}(\kappa) \left( -1 + z^{-\kappa} \right)  \right]^{1/2} dz &= \left( \frac{\pi}{2}\right)^{1/2} \frac{\displaystyle \Gamma\left[ \frac{\left( \mathscr{D}_{\text{v}} - \mathscr{D}_{\text{s}} \right)}{\kappa} - \frac12 \right]}{\displaystyle \kappa \Gamma\left[ \frac{\left( \mathscr{D}_{\text{v}} - \mathscr{D}_{\text{s}} \right)}{\kappa} + 1 \right]} \ \ \text{for} \ \ \kappa > 0 \ \ ,
\\
&= \left( \frac{\pi}{2}\right)^{1/2} \frac{\displaystyle \Gamma\left[ \frac{\left( \mathscr{D}_{\text{v}} - \mathscr{D}_{\text{s}} \right)}{|\kappa|} \right]}{\displaystyle |\kappa| \Gamma\left[ \frac{\left( \mathscr{D}_{\text{v}} - \mathscr{D}_{\text{s}} \right)}{|\kappa|} + \frac32 \right]} \ \ \text{for} \ \ \kappa < 0 \ \ ,
\end{split}
\end{equation}
and hence we obtain the Rydberg sizes and energies as follows:
\begin{equation}
\tilde{r}_{\max} = \left( \frac{\pi n}{\Theta} \right)^{\frac{2}{2\left(\mathscr{D}_{\text{v}}-\mathscr{D}_{\text{s}} \right)-\kappa}} \ \ , \ \ \left| \tilde{E} \right| = \left( \frac{\pi n}{\Theta} \right)^{-\frac{2 \kappa}{2\left(\mathscr{D}_{\text{v}}-\mathscr{D}_{\text{s}} \right)-\kappa}}  \ \ .
\label{R_E_asymp}
\end{equation}

\ \ 

\bibliography{main}
\bibliographystyle{apsrev4-2}

\end{document}